\documentclass[pra,twocolumn,superscriptaddress,showpacs]{revtex4-1}
\usepackage[english]{babel}
\usepackage{graphicx}
\usepackage{amsmath}
\usepackage{amsfonts}
\usepackage{bm}
\usepackage{color}
\usepackage{epstopdf}

\newcommand{\ket}[1]{|#1\rangle}

\definecolor{darkgreen}{rgb}{0,0.60,.2}

\begin{document}
\title{Mechanism of Ultrafast Relaxation of a Photo-Carrier in Antiferromagnetic Spin Background}

\author{Denis \surname{Gole\v{z}}}
\affiliation{J. Stefan Institute, 1000 Ljubljana, Slovenia}

\author{Janez \surname{Bon\v{c}a}}
\affiliation{J. Stefan Institute, 1000 Ljubljana, Slovenia}
\affiliation{Faculty of Mathematics and Physics, University of Ljubljana, 1000
Ljubljana, Slovenia}

\author{Marcin \surname{Mierzejewski}}
\affiliation{Institute of Physics, University of Silesia, 40-007 Katowice, Poland}

\author{Lev \surname{Vidmar}}
\affiliation{Department of Physics and Arnold Sommerfeld Center for Theoretical Physics, Ludwig-Maximilians-Universit\"at M\"unchen, D-80333 M\"unchen, Germany}
\affiliation{J. Stefan Institute, 1000 Ljubljana, Slovenia}


\begin{abstract}
We study the relaxation mechanism of a highly excited charge carrier propagating in the antiferromagnetic background modeled by the $t$-$J$ Hamiltonian on a square lattice.
We show that the relaxation consists of two distinct stages.
The initial ultrafast stage with the relaxation time $\tau$$\sim$$(\hbar/t_0)(J/t_0)^{-2/3}$ (where $t_0$ is the hopping integral and $J$ is the exchange interaction) is based on generation of string states in the close proximity of the carrier.
This  unusual scaling of $\tau$  is obtained by means of  comparison of numerical results with a simplified $t$-$J_z$ model on a Bethe lattice.
In the subsequent (much slower) stage  local antiferromagnetic excitations are carried  away by magnons.
The relaxation time on the two-leg ladder system is an order of magnitude longer due to the lack of string excitations.
This further reinforces the importance of string excitations for the ultrafast relaxation in the two-dimensional  system. 
\end{abstract}
\maketitle

In a large number of generic time-dependent many-body systems, it is conjectured that strong electronic correlations give rise to extremely fast timescales of relevant relaxation processes.
In general, the nonequlibrium evolution of a simple quantum system is a complex problem with only a few exactly solvable cases, and strong interactions between charge carriers usually make the problem even harder.
Although advanced numerical approaches give important information about nonequlibrium dynamics, this dynamics is in many cases too complex to be comprehended in terms of a simple physical picture.
Distinguishing different elementary excitations in time domain therefore represents one of the major goals of the present research of nonequilibirum many-body systems.
In this context, a rapid development of time-resolved experiments in condensed-matter systems~\cite{dalconte12,fausti11,gadermaier10,*gadermaier12,cortes2011,*rettig12,kim12,novelli12,mansart13} and cold atomic gases~\cite{bloch08} provide both a challenge for theory as well as a testbed for new ideas.
A large body of current theoretical research is based on studies of Hubbard-like models far from equilibrium, and
it focuses both on relaxation dynamics after a sudden quench~\cite{moeckel08,eckstein09,schiro10,*schiro11,hamerla13a,*hamerla14,stark13} and steady-state properties as a consequence of constant driving~\cite{oka2003,*oka2005,joura08,mierzejewski2010,*mierzejewski2011a,prosen2010,eckstein2010,*zala2012,*aron12,heidrichmeisner10,*busser13,kirino10,eckstein11b,aron2012,*amaricci11,einhellinger12,bukov2012,alvermann12,han13,*han13b}.

Understanding the dynamics of photo-induced charge carriers in Mott insulators may contribute  to unravelling still elusive mechanism of high-$T_C$ superconductivity, in addition,  it   is as well indispensable   for applications of novel materials in future electronic and photovoltaic technologies.
Many recent studies focused on dynamics of photo-induced carriers, {\it i.e.}, doublons and holons~\cite{mitrano13,takahashi2002,koshibae09,kanamori09,freericks09,moritz10,matsueda11,defilippis12,rademaker12a,*rademaker12b,eckstein2012,al-Hassanieh2008,dias2012},
in particular on their nonradiative recombination process~\cite{strohmaier2010,*sensarma2010,eckstein11a,hofmann2012,al-Hassanieh2008,dias2012,zala2012a,andre12}.
Experimentally, photo-induced carriers have been observed in, e.g., insulating cuprates, where they decay within a few hundreds of femtoseconds~\cite{okamoto2010,*okamoto2011}.

In this manuscript, we investigate a phenomenon which precedes the recombination of photo-induced carriers, {\it i.e.}, we consider a rapid exchange of energy between photo-induced carriers and their local environment.
In fact, the study does not only address Mott insulators where all the charge carriers are photo-{\it induced}, but also doped Mott insulators where doped charge carriers are photo-{\it excited}.
We therefore apply the term photo-carriers throughout the manuscript to generally describe the highly excited charge degrees of freedom.
We base our study on the following main assumptions:
{\it (i)}
Kinetic energy of photo-carriers is instantly raised to values much larger than their equilibrium value;
{\it (ii)}
Photo-carriers propagate in a background with at least short-range antiferromagnetic (AFM) correlations;
{\it (iii)}
We only consider energy transfer from charge to spin degrees of freedom.
We propose a microscopic mechanism of extremely fast primary relaxation of photo-carriers propagating in the AFM background on a square lattice.
The mechanism is based on a generation of local AFM excitations (denoted also as string states) in the close proximity of the photo-carrier.
We argue that the essential physics of the primary relaxation can be accurately described by a simple model on a Bethe lattice.
At a later time, secondary relaxation process describes the dissipation of the local excess magnetic energy via magnons.

\begin{figure*}[!tb]
\includegraphics[width=0.90\textwidth]{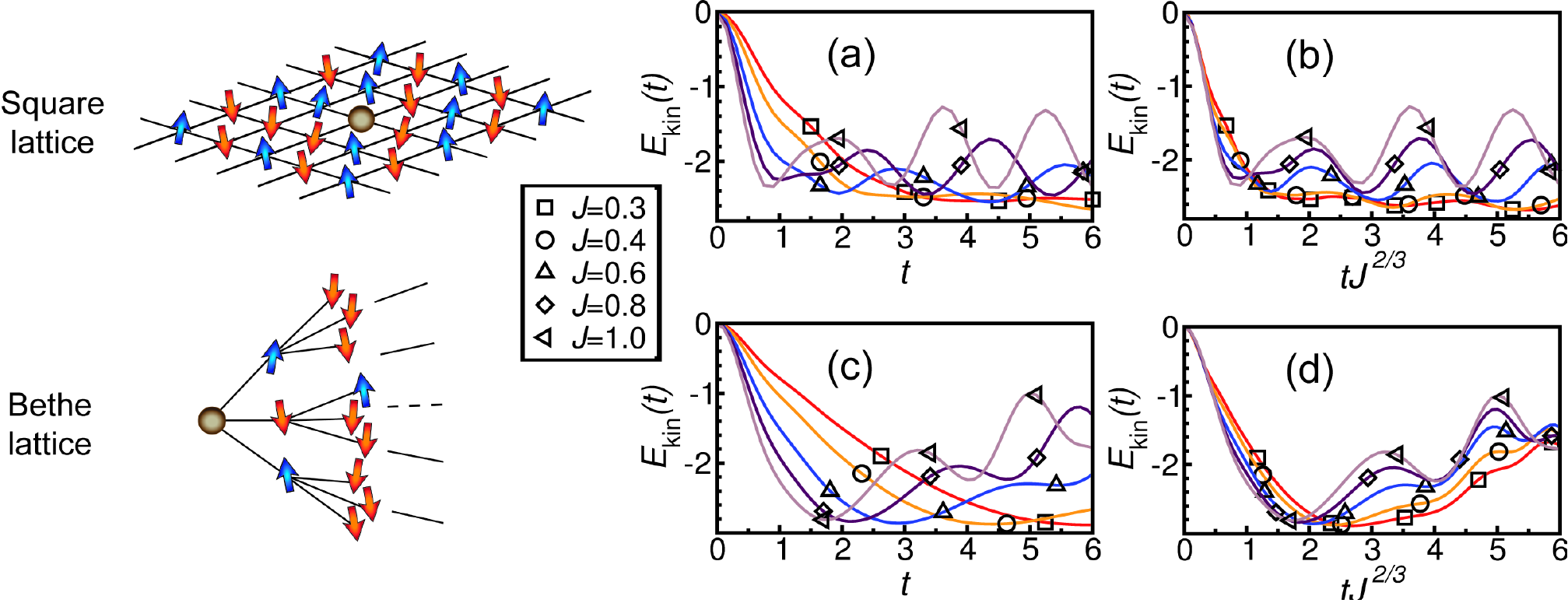}
\caption{
{\it Relaxation dynamics: square lattice vs Bethe lattice.}
Square lattice, $t$-$J$ model:
$E_{\rm kin}(t)$ after the phase quench $\Delta \phi_{\bf i,i+e_x}$=$\pi$, as a function of (a) time $t$ and (b) rescaled time $t$$\to$$tJ^{2/3}$.
Bethe lattice, $t$-$J_z$ model, Eq.~(\ref{hambet}):
$E_{\rm kin}(t)$ for the hopping amplitude quench (see Appenidx~\ref{s1}), as a function of (c) time $t$ and (d) rescaled time $t$$\to$$tJ^{2/3}$ (we used $J$=$J_z$ in lower panels).
}
\label{fig1}
\end{figure*}

As discussed in Ref.~\cite{al-Hassanieh2008}, absorption of photons by solids is a complex process which initially evolves through states not included in tight-binding Hamiltonians. 
Due to multiple scattering events the photo-carriers quickly dissipate their energy and enter the regime which is captured by the tight-binding model.
Then, however, photo-carriers are inserted with arbitrary (typically large) kinetic energy. 
We model such situation by considering a charge carrier within the $t$-$J$ model on a square lattice,
\begin{eqnarray}
 H&=&H_\mathrm{kin}+H_J \label{ham}  \\
    &=& -t_{0}\sum_{\langle\bf{i,j}\rangle,\sigma} [e^{i\phi_{\bf i,j}(t)} \tilde c_{\bf{i},\sigma}^{\dagger} \tilde c_{\bf{j},\sigma} + \mbox{H.c.} ] + J\sum_{\langle\bf{i,j}\rangle}\mathbf{S}_{\bf i} \cdot \mathbf{S}_{\bf j}, \nonumber
\end{eqnarray}
where $\tilde c_{{\bf i},\sigma}$=$c_{{\bf i},\sigma}(1$-$n_{{\bf i},-\sigma})$ is a projected fermion operator and $\langle\bf{i,j}\rangle$ denotes nearest neighbors.
The system is threaded by a time-dependent flux, which induces the electric field $ - \partial_t {\phi}_{\bf i,j}(t)$.
Hence the $\delta$-like pulse of the electric field can be described by the sudden increase (quench) of $\phi_{\bf i,j}$.
After calculating the ground state of the model with $\phi_{\bf i,j}$=$0$, we apply at time $t$=$0$ 
a {\it phase quench} setting  $\phi_{\bf i,i+e_x}(t)$=$\pi \theta(t)$, where $\bf e_x$ represents the unit vector in the $x$-direction.
The effect of this particular quench is to change the sign of hopping in the $x$-direction that  consequently leads to a sudden increase of the kinetic energy to $E_\mathrm{kin}(t$=$0)$=$\langle H_\mathrm{kin}(t$=$0)\rangle$=$0$.
In the Appendix~\ref{s1} we study other types of quenches and show that the main results of the study are independent of the particular form of the quench.
We used the dimensionless units by putting $\hbar$=$t_0$=$1$.

We employ diagonalization in a limited functional space (see Appendix~\ref{s0}), which was successfully used to describe equilibrium and nonequilibrium properties of a charge carrier doped into a planar ordered antiferromagnet~\cite{bonca2007,mierzejewski2011,bonca2012}.
Applying the Lanczos technique  we first compute  the initial ground state $|\Psi(t$=$0)\rangle$, then we  implement  the  time evolution
$\ket{\Psi(t)}$=$e^{-iHt} |\Psi(t$=$0)\rangle$
using the quenched Hamiltonian.
At each small time step $\delta t$$\ll$$1$ we use Lanczos basis for generating the evolution $|\Psi(t-\delta t)\rangle$$\rightarrow$$\ket{\Psi(t)}$~\cite{mierzejewski2010,mierzejewski2011,park1986}.


In Figure~\ref{fig1}(a) we present the time evolution of the kinetic energy of a photo-carrier after the phase quench.
We observe the ultrafast relaxation shortly after the quench, as $E_\mathrm{kin}(t)$ rapidly decreases from its initial value $E_\mathrm{kin}(t$=$0)$=$0$ towards the pre-quench value.
We denote this process a primary relaxation. 
In addition, we observe that the characteristic relaxation time decreases with increasing $J$.
Figure~\ref{fig1}(b) reveals seemingly unusual scaling with time $t$$\to$$tJ^{2/3}$ that turns out to be nearly perfect for all values of $J$ at short rescaled times, {\it i.e.,} for $tJ^{2/3}$$\lesssim$$0.4$.

A closer inspection in the physically relevant regime of small $J$$\leq$$0.4$ and times  $tJ^{2/3}$$\gtrsim$$1.0$ suggests a secondary, much longer relaxation time slightly masked by the superimposed oscillations.
We address this issue via a simple analytical scenario  towards the end of the paper.

\begin{figure*}[!t]
\includegraphics[width=0.90\textwidth]{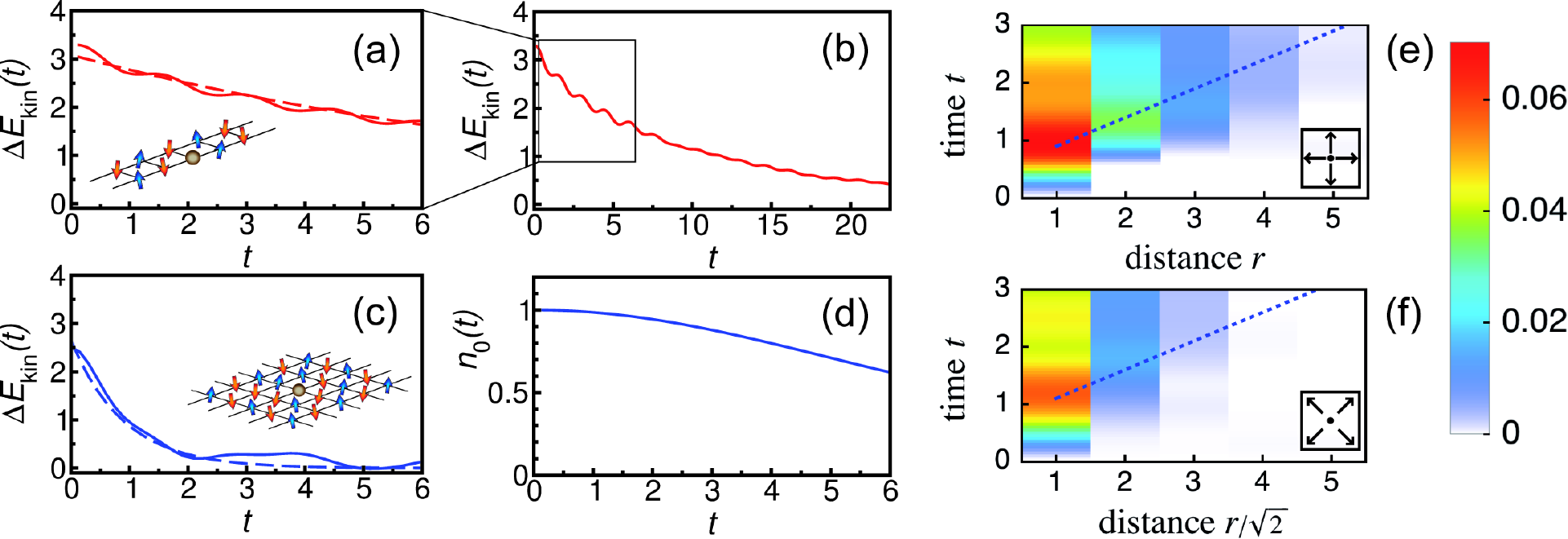}
\caption{
{\it Relaxation dynamics: two-leg ladder vs square lattice.}
(a) and (b) $\Delta E_\mathrm{kin}(t)$=$E_\mathrm{kin}(t)$-$E_\mathrm{kin}(t$$\to$$\infty)$ on a two-leg ladder, (c) $\Delta E_\mathrm{kin}(t)$ on a square lattice.
On a ladder we use exact diagonalization on 24 sites with periodic boundary conditions.
%
Dashed lines represent exponential fits
$\Delta E_\mathrm{kin}(t)$=$A \exp{(-t/\tau)}$.
We get $A_{\rm ladd}$=$3.1$ and $\tau_{\rm ladd}$=$9.5$ in (a) and $A$=$2.6$ and $\tau$=$0.9$ in (c).
In (d) we present the occupancy of a free particle $n_0(t)$ with magnon dispersion on a square lattice, localized at site 0 at $t$=$0$.
(e) and (f)
The correlation function on a square lattice
$C(|\mathbf{r}|,t)$=$\sum_{\mathbf{i}} | \langle (1$-$n_{\mathbf{i}}) (S_{\mathbf{i}+\mathbf{r}}^{z,\mbox{\tiny N\' eel}}$-$S_{\mathbf{i}+\mathbf{r}}^{z})  \rangle|/ N(|{\bf r}|)$
as a measure of the deviation from the N\' eel   state
$S_{\mathbf{j}}^{z,\mbox{\tiny N\' eel}}$=$\pm \frac{1}{2}$, 
where $n_{\bf i}$=$\sum_\sigma n_{{\bf i},\sigma}$ and $N(|{\bf r}|)$ denotes the number of sites at distance $r$=$|{\bf r}|$.
The density plots in (e) and (f) represent
$C(\tilde r,t)$-$C(\tilde r,t$=$0)$ along the $x$- and $y$-axis ($\tilde r$=$r$) and along the diagonals ($\tilde r$=$r/\sqrt{2}$), respectively.
The dashed lines $\tilde r$=$v_0 t$+$\delta$ indicate the spreading given by the group velocity $v_0$=$2t_0$.
In all the panels, we set $J$=$0.6$ and measure time in units of $[1/t_0]$.
}
\label{fig2}
\end{figure*}


The short-time scaling with $t$$\to$$tJ^{2/3}$ observed in Fig.~\ref{fig1}(b) suggests that the key ingredient of the fast relaxation mechanism is the generation of local AFM excitations by the photo-carrier.
These excitations are also known as string states~\cite{bulaevskii68,shraiman1988,*dagotto1994}.
We now present a simple model of the ultrafast relaxation mechanism via string states which explains the scaling found in Fig.~\ref{fig1}(b).
We use the $t$-$J_z$ model~\footnote{The $t$-$J_z$ model is obtained from the $t$-$J$ model, Eq.~(\ref{ham}), by taking into account only the $z$-components of spin operators.}
on a Bethe lattice with connectivity $z$=$4$.
An effective Hamiltonian $H'$ for a single charge carrier (hole) in a N\'{e}el state is written within the string basis $\vert l \rangle$ representing strings with a given length $l$,
\begin{align}
&H'\ket{0} = J_z \ket{0}-2t_0\ket{1}, \nonumber \\
&H' \ket{1}= J_z {5\over 2}\ket{1}-2t_0\ket{0}-\sqrt{3}t_0\ket{2},\label{hambet} \\
&H' \ket{l}= J_z ({3\over 2}+l)\ket{l}-\sqrt{3}t_0(\ket{l-1}+\ket{l+1}); \; l \geq 2, \nonumber
\end{align}
where the energy is measured from the N\'{e}el state with no carriers~\cite{shraiman1988,*dagotto1994}.
For clarity we have reintroduced the original units.
In the continuum limit $| l$+$1\rangle$=$(1$+$a\frac{d}{dl}$+$\frac{1}{2} a^2\frac{d^2}{dl^2})\ket{l} $,  where $a$ is the lattice distance, the problem in Eq.~(\ref{hambet}) reduces for $l$$>$$1$ to a particle in a linear potential.
The corresponding time-dependent  Schr\"odinger equation 
$i\frac{d \phi}{dT}$=$[-\sqrt{3}\frac{d^2}{dx^2}$$+$$x]$$\phi$
can be obtained by introducing the dimensionless variables $x$=$(J_z/t_0)^{1/3}l/a$ and $T$=$t (t_0/\hbar) (J_{z}/t_0)^{2/3}$.

Note that Eq. (\ref{hambet}) as a tridiagonal eigenproblem technically does not pose any serious computational challenge.
Our objective is to show whether/when this simplified approach describes the relaxation in the physical charge-spin systems like the $t$-$J$ model.
Below we show that the Hamiltonian (\ref{hambet}) is indeed relevant for the initial stage of the relaxation.
While it does not allow for a direct description of a {\it phase quench}, it is well suited for the investigation of a so-called {\it polaron formation} process~\cite{ku2007,fehske11}, {\it i.e.}, starting from a state of a carrier in the N\'{e}el background with no string states, described by the state $\ket{0}$.
This is achieved by quenching the amplitude of the hopping integral $t_0(t)$=$t_0\theta(t)$, leading to the same initial expectation value $\langle 0\vert H_\mathrm{kin}\vert 0 \rangle$=$0$ as in the case of the phase quench (we further compare different quench protocols in Appendix~\ref{s1}).
In Fig.~\ref{fig1}(c) we observe an initial fast decrease of $E_\mathrm{kin}(t)$ as well as a decreasing relaxation time with increasing $J_z$.
Figure~\ref{fig1}(d) shows the time scaling $t$$\to$$tJ_z^{2/3}$.
There is a remarkable agreement between the $t$-$J$ model on a square lattice and the $t$-$J_z$ model on a Bethe lattice regarding the primary relaxation process at $tJ^{2/3}$$\lesssim$$1$, see Figs.~\ref{fig1}(b) and~\ref{fig1}(d)
[note that a perfect scaling of curves with different values of $J$ only exists within the continuous model derived by neglecting peculiarities at $l$=$0$ and $1$ in Eq.~(\ref{hambet})].
At longer times, the most prominent difference is a partial recovery of $E_\mathrm{kin}(t)$ in Fig.~\ref{fig1}(d) that appears as a  consequence  of a weaker damping in the simplified model of  Eq.~(\ref{hambet}).
Stronger damping  found in the $t$-$J$ model is attributed in part to quantum antiferromagnetic fluctuations as well as to a higher complexity of the functional space spanning the latter model.

While the main goal of the present study is to focus on a qualitative description of the relaxation dynamics, we can as well make some quantitative estimates of relaxation times.
Even though the relaxation dynamics at very short times does not resemble an exponential decay, a rough  estimate of the characteristic relaxation time from an exponential fit in the regime $J$$\leq$$0.4$ of Fig.~\ref{fig1}(b) yields
$\tau$$\sim$$0.8J^{-2/3}$.
By using the original time-units and model parameters relevant for materials such as cuprate superconductors, {\it i.e.}, $t_0$=$0.4$eV and $J/t_0$=$0.3$, this gives
$\tau$$\sim$$0.8(\hbar/t_0)(J/t_0)^{-2/3}$$\sim$$3.0 \mbox{fs}$.


Since the initial kinetic energy of the photo-carrier strongly exceeds the energy that a single AFM bond can  accommodate, we may intuitively expect that a large AFM reservoir in the form of multiple  configurations of different string states in close proximity of the photo-carrier may absorb the energy more efficiently. 
We test this conjecture by comparing relaxation in the $t$-$J$ model on a square lattice and a two-leg ladder.
Figures~\ref{fig2}(a)-\ref{fig2}(c) present the relaxation of kinetic energy after the phase quench $\phi_{\bf i,i+e_x}(t)$=$\pi \theta(t)$.
We observe an exponential decay on a ladder system,
see Fig.~\ref{fig2}(b).
The comparison of fits from Figs.~\ref{fig2}(a) and~\ref{fig2}(c) reveal that the relaxation time on a two-leg ladder is an order of magnitude longer than on a square lattice,
$\tau_{\rm ladd}/\tau$$\approx$$10$.
A much slower relaxation of the quasi-one-dimensional ladder system is consistent with studies of one-dimensional Hubbard models~\cite{al-Hassanieh2008,matsueda11}, where even longer relaxation times were observed.
This result further reinforces the relaxation mechanism  based on  generation of local string states that leads to an unusually fast primary relaxation in the two-dimensional system.

While the phase quench instantaneously  increases the kinetic energy of the photo-carrier, it does not directly affect the AFM background.
In Figs.~\ref{fig2}(e) and~\ref{fig2}(f) we plot the correlation function showing the time-evolution of the AFM background on a square lattice.
Even though the fastest spread of AFM excitations is roughly given by the maximal group velocity of the carrier, it confirms our expectations that in the primary relaxation, the strongest perturbation of the AFM background is limited to the close vicinity of the photo-carrier.
Since this mechanism requires only short-range AFM order, it should be efficient far beyond the boundaries of the long-range-ordered AFM phase.
Experimental detection of short-range AFM correlations represents a timely research topic~\cite{tacon11,dean13} and indicates that the coupling to these excitations can eventually be detected in pump-probe experiments with high time-resolution.

Up to this moment we have focused on the primary relaxation in the ultrafast regime $tJ^{2/3}$$\lesssim$$1$.
Comparison between the $t$-$J$ model on a square lattice and the $t$-$J_z$ model on a Bethe lattice, Figs.~\ref{fig1}(b) and~\ref{fig1}(d), undisputedly suggest that in this regime magnons do not play any significant role since the photo-carrier releases its excess kinetic energy to local AFM excitations.
This is consistent with the observation that for small and moderate values of $J$, the characteristic time of primary relaxation is noticeably faster than the dynamics of AFM excitations (magnons) in the $t$-$J$ model, $J^{-2/3}$$<$$J^{-1}$.
In order to provide an analytical evidence for this scenario we consider a noninteracting particle propagating on a square lattice with $N$ sites and with the same dispersion as magnons:
$\omega_{\mathbf k}$=$\frac{1}{4}\sqrt{J^2(0)-J^2(\mathbf{k})}$,
where  
$J(\mathbf{k})$=$2J(\cos k_x$+$\cos k_y)$.
We assume that such a particle is created at the position $\mathbf{R}$=$0$ at time $t$=$0$.
Then, the average occupation of site $\mathbf{R}$=$0$ changes in time according to
$n_0(t)$=$|\frac{1}{N}\sum_{\mathbf{k}} \exp(it\omega_\mathbf{k})|^2$, as shown in Fig.~\ref{fig2}(d).
Comparison to $E_{\rm kin}(t)$ in Fig.~\ref{fig2}(c) reveals that $n_{0}(t)$ retains its initial value within the entire time window where the primary relaxation of $E_{\rm kin}(t)$ takes place.
The antiferromagnetic dynamics as a secondary relaxation process in the $t$-$J$ model is therefore fairly disentangled from the primary relaxation.

In conclusion, we have proposed a microscopic mechanism of an ultrafast energy transfer of the order of 1eV from a charge carrier to AFM excitations on a few-femtosecond time scale.
In this primary relaxation stage a photo-carrier creates local string states in the magnetic background.
The relaxation is extremely fast not only due to the high energy of the relevant AFM excitations, but because the photo-carrier is inherently strongly coupled to the AFM background. 
While the present study accounts for the existence of long-range AFM correlations, the key ingredients of the ultrafast relaxation are nevertheless local AFM excitations in the proximity of the photo-carrier.
It is therefore very likely that the presented mechanism acts as an important relaxation channel also in systems with finite doping and short-range AFM correlations.
One needs to keep in mind, though, that a quantitative comparison to the later systems is not possible since in the current calculation we include, in principle, an infinite reservoir of AFM excitations per single photo-carrier.
Since the main effect of finite doing on the relaxation time would emerge through the decrease of the available AFM excitations per doped carrier, we expect the increase of the relaxation time with doping.

The primary relaxation leads to a strongly non-thermal local states of the AFM background.
In the $t$-$J$ model, subsequent thermalization of the AFM background through propagating magnons represents the secondary, usually much slower stage of the relaxation.
Disentanglement and characterization of the two different relaxation processes within the $t$-$J$ model represents an important step in understanding more complex systems where competing interactions, e.g., phonons, may also represent an efficient relaxation channel~\cite{kabanov2008,*baranov13,vidmar11}.
The two-stage relaxation mechanism appears to be in agreement with studies on the Hubbard-like small clusters~\cite{takahashi2002}.
Note however, that in the latter study the primary relaxation time does not seem to be of the magnetic origin since a scaling with $\tau$$\sim$$\hbar/t_{0}$ is proposed, while our calculations yield $\tau$$\sim$$(\hbar/t_0)(J/t_0)^{-2/3}$, which is a clear sign of   magnetic origin.

Recently, two studies appeared on the preprint server that address a very similar topic~\cite{iyoda13,*eckstein14}.
We acknowledge stimulating discussions with
U. Bovensiepen, S. Dal Conte, M. Eckstein, C. Gadermaier, C. Giannetti, Z. Lenar\v ci\v c, D. Mihailovi\'{c} and P. Prelov\v sek.
J. B. acknowledges support by the P1-0044 of ARRS, Slovenia.
M. M. acknowledges support from the NCN project DEC-2013/09/B/ST3/01659.
L.V. is supported by Alexander von Humboldt Foundation.
This work was performed, in part, at the Center for Integrated Nanotechnologies, an Office of Science User Facility operated for the U.S.

\appendix

\section{Diagonalization in a limited functional space} \label{s0}

In this Appendix we provide details about the functional space that we use to calculate the dynamics on the square lattice.
We employ diagonalization in a limited functional space to calculate both the initial wavefunction of a single carrier in the $t$-$J$ model on a square lattice, as well as the relaxation dynamics after a sudden quench.
The generation of the functional space starts from a translationally invariant state of a carrier in the N\' eel background $|\phi_0\rangle=c_{\mathbf{k}}|\mbox{N\' eel}\rangle$ with $\mathbf{k}=(\pi/2,\pi/2)$.
The kinetic part $H_{\rm kin}$ as well as the off-diagonal spin-flip part $\tilde{H}_J$ of the time-independent $t$-$J$ Hamiltonian, Eq.~(1), are applied up to $N_h$ times generating the basis functions
\begin{equation} \label{lfs}
\left\{|\phi_{l}^{n_h} \rangle \right\}=[H_{\rm kin}(\phi_{\bf i,j}=0) + \tilde{H}_J]^{n_h} |\phi_0 \rangle
\end{equation}
for $n_h=0,...,N_h$.
During the generation of states only translationally invariant parent states are kept.
Ground-state properties of the system obtained by diagonalizing the functional space, Eq.~(\ref{lfs}), showed perfect agreement with other methods developed to study properties of a single carrier in the $t$-$J$ model~\cite{bonca2007}.

\begin{figure}[!tb]
\includegraphics[width=0.49\textwidth]{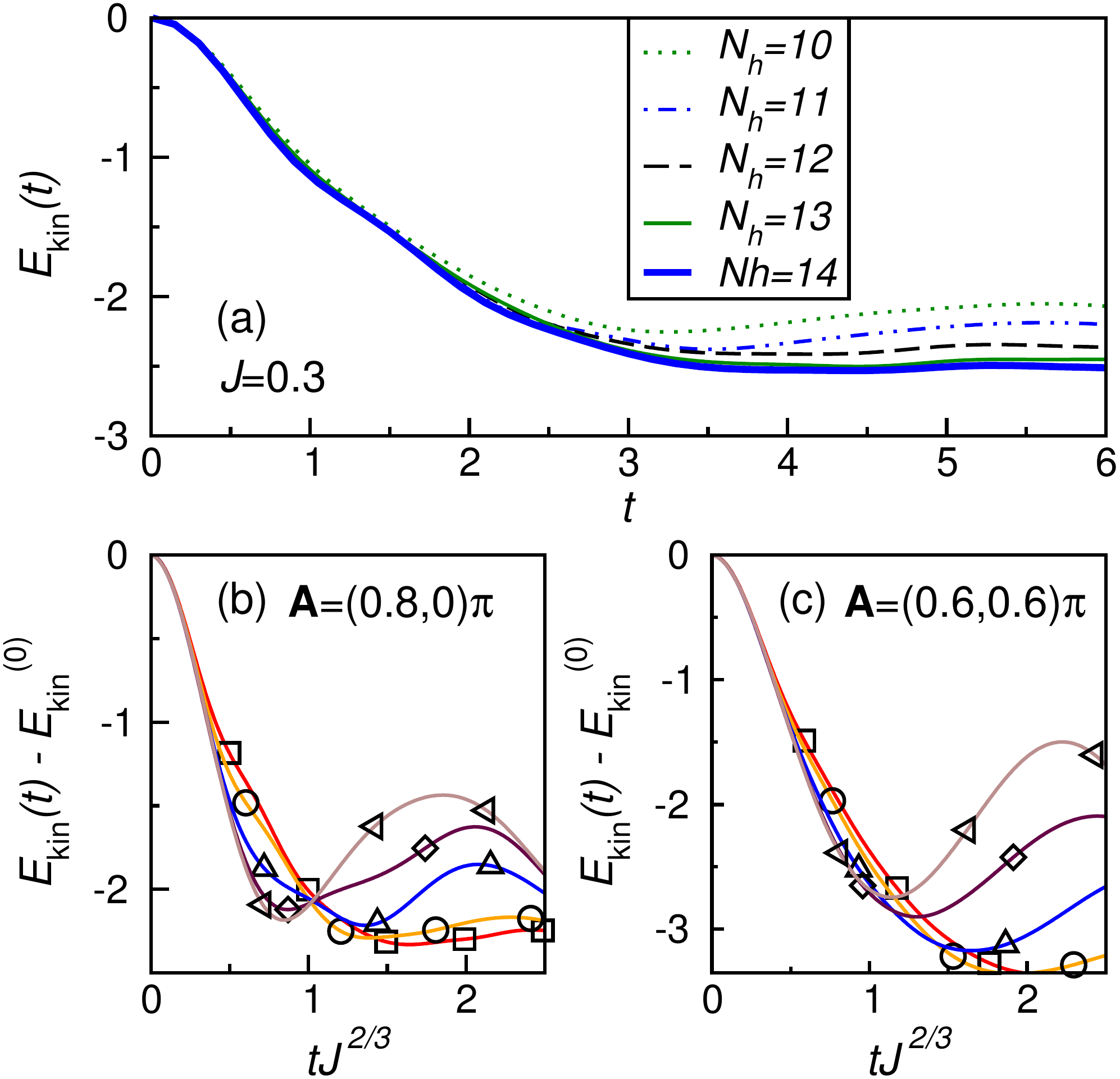}
\caption{
{\it Relaxation dynamics on a square lattice.}
(a)
$E_\mathrm{kin}(t)$ for $J=0.3$ and the phase quench $\phi_{\bf i,i+e_{x(y)}}(t)$=$A_{x(y)} \theta(t)$, where $A_x$=$\pi$ and $A_y$=$0$.
Five nearly overlapping curves represent results for different values of $N_h=10,...,14$ with functional spaces ranging from $N_{\rm states}=1.7 \times 10^5$ up to $N_{\rm states}=1.7 \times 10^7$.
(b) and (c)
$E_\mathrm{kin}(t)$ for the phase quench ${\bf A}=(0.8,0)\pi$ and ${\bf A}=(0.6,0.6)\pi$, respectively.
We use different values of $J$, {\it i.e.}, $J=0.3$ (squares), $J=0.4$ (circles), $J=0.6$ (triangles up), $J=0.8$ (diamonds) and $J=1.0$ (triangles left).
We subtract the curves by the initial values of the kinetic energy $E_{\rm kin}^{(0)}$ instantly after the quench, such that all the curves start from the same initial value.
}
\label{Fig:sup0}
\end{figure}

After quenching the Hamiltonian we calculate the time-evolution of the wavefunction within the same functional space.
The advantage of the diagonalization in the limited functional space over the standard exact diagonalization follows from a systematic generation of selected states which contain local antiferromagnetic excitations in the vicinity of the carrier.
As a consequence, it enables the investigation of the dynamics of large systems, which are far beyond the reach of exact diagonalization.
The diagonalization in the limited functional space was successfully applied to calculate steady state properties of a single carrier~\cite{mierzejewski2011} and two charge carriers~\cite{bonca2012} in the $t$-$J$ model driven by a constant electric field.
The parameter $N_h$ determines the accessible energies (the number of local antiferromagnetic excitations) and should be large enough such that the relaxation dynamics is independent of $N_h$.
We set $N_h=14$ in all our calculations.
In Fig.~\ref{Fig:sup0}(a) we compare the kinetic energy $E_{\rm kin}(t)$ of the charge carrier for different values of parameter $N_h$.
Remarkably, even though the sizes of the functional space shown in the figure extend over two orders of magnitude, different curves representing $E_{\rm kin}(t)$ in Fig.~\ref{Fig:sup0}(a) are virtually indistinguishable in the time interval $t \leq 3$.
The core reason for the excellent convergence stems from the local relaxation mechanism discussed in the paper, {\it i.e.}, the photo-carrier creates local antiferromagnetic excitations in its close vicinity.
As long as the local antiferromagnetic excitations do not propagate over large distances, the functional space includes the relevant spin states.
This explains the efficiency of the method in the nonequilibrium regime after the quench.
The results therefore reveal that the ultrafast primary relaxation of the photo-carrier in the $t$-$J$ model barely depends on the size of functional space provided $N_h > 10$.
The main limitation of the method represents the time-window in which converged results are obtained, see the largest times in Fig.~\ref{Fig:sup0}(a).
This effect becomes important when the antiferromagnetic excitations start propagating over large distances.

\section{Phase quench versus hopping amplitude quench} \label{s1}

We next compare different quench protocols to suddenly increase the kinetic energy of a photo-carrier.
In the main part of the paper, we used the following quench protocol:
We first calculated the ground state of a charge carrier in the $t$-$J$ model, then we suddenly increased the carrier's kinetic energy by applying the phase quench $\phi_{\bf i,i+e_{x(y)}}(t)$=$A_{x(y)} \theta(t)$, where $A_x=\pi$, $A_y=0$ and $\bf e_{x(y)}$ represents the unit vector in the  $x(y)$-direction.
The reason for this particular choice of the phase quench with ${\bf A}=(\pi,0)$ is to relate the time evolution of the kinetic energy $E_\mathrm{kin}(t)$ to the $t$-$J_z$ model on the Bethe lattice, where the hopping amplitude was quenched.
In both cases (phase quench with ${\bf A}=(\pi,0)$ and the hopping amplitude quench) the initial kinetic energies were $E_{\rm kin}(t=0)=0$, which  allowed for a direct comparison of results. 
Nevertheless, we show in the following that the observed scaling of the relaxation time $\tau \sim (\hbar/t_0)(J/t_0)^{-2/3}$ is independent of the particular value of the phase quench ${\bf A}=(\pi,0)$.
We show in Fig.~\ref{Fig:sup0}(b) and (c) the relaxation of $E_\mathrm{kin}(t)$ for ${\bf A}=(0.8,0)\pi$ and ${\bf A}=(0.6,0.6)\pi$, respectively, for different values of $J$.
In both cases we observe a universal relaxation of $E_\mathrm{kin}$ as a function of $tJ^{2/3}$ in the time interval $tJ^{2/3} \lesssim 0.5$.
Remarkably, in the latter time regime, nearly half of the kinetic energy is already transferred to local antiferromagnetic excitations.

In addition, we studied the $t$-$J_z$ model on the Bethe lattice, where the initial wavefunction was the N\' eel state with a single localized carrier.
Hence a quench of the hopping amplitude $t_0(t) = t_0 \theta(t)$ was performed.
The time-evolution of such system is also denoted as the {\it polaron formation} process.
The equivalent set-up for the $t$-$J$ model on a square lattice is to calculate the ground state of the Heisenberg model, $ \ket{\psi_{\rm AFM}}$, and then to replace one spin by a charge carrier (hole) in a translationally invariant state with $\mathbf{k}=(\pi/2,\pi/2)$,
\begin{equation} \label{def_polform}
 \ket{\psi_0} = c_{\bf k} \ket{\psi_{\rm AFM}}. 
\end{equation}
We show in this Appendix that different quench protocols in the $t$-$J$ model, {\it i.e.}, the phase quench and the hopping amplitude quench, lead to the same qualitative description of the ultrafast relaxation dynamics.

\begin{figure}[!tb]
\includegraphics[width=0.49\textwidth]{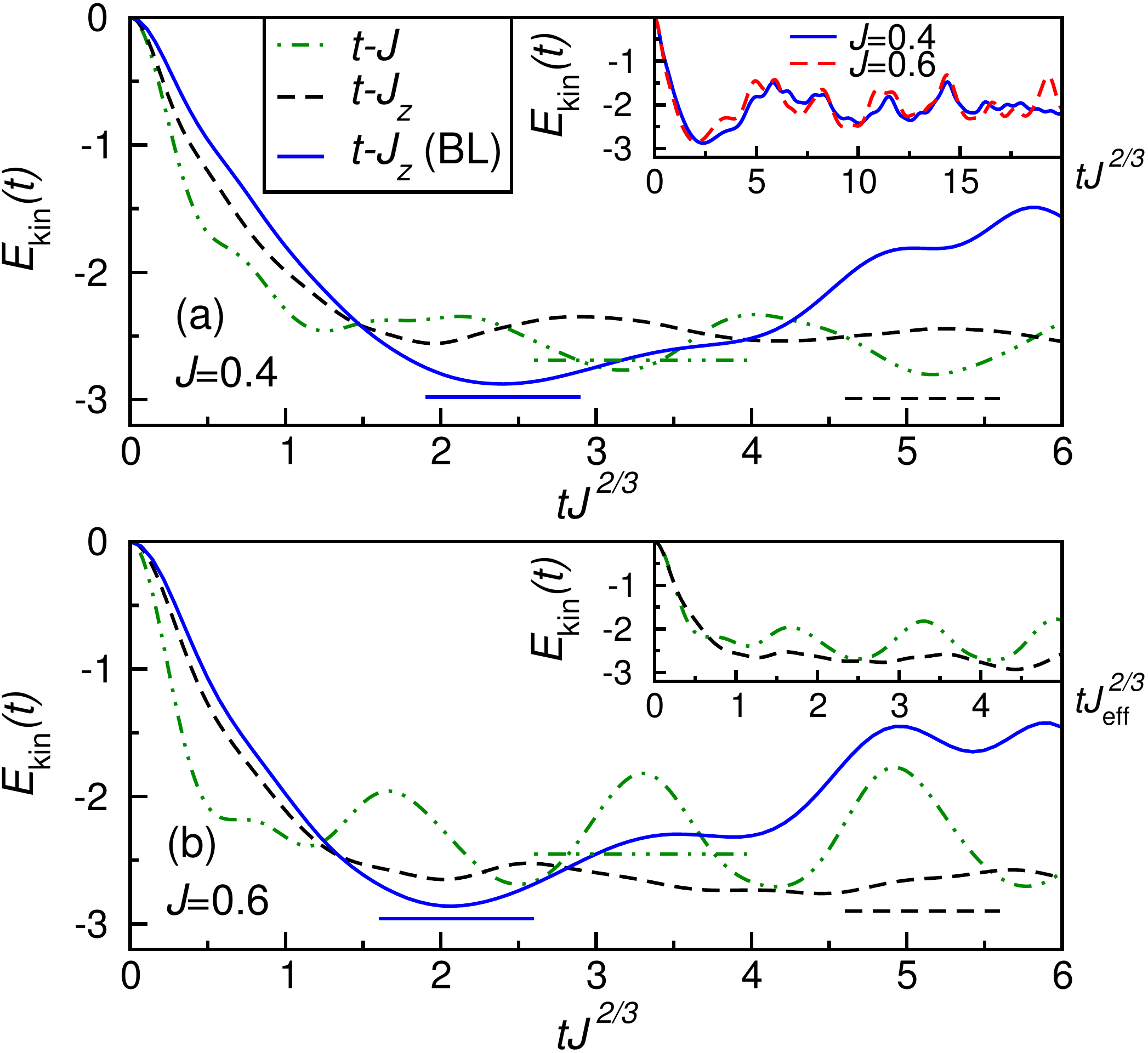}
\caption{
{\it Comparison of the polaron formation in the $t$-$J$ model, the $t$-$J_z$ model (both on a square lattice, using diagonalization within the functional space defined in Eq.~(\ref{lfs})), and the $t$-$J_z$ model on the Bethe lattice (BL).}
We plot the kinetic energy of the charge carrier $E_{\rm kin}$ vs rescaled time $tJ^{2/3}$ for different exchange interactions $J\equiv J_z=0.4$ in (a) and $0.6$ in (b).
Inset in (a) shows results of the $t$-$J_z$ model on the Bethe lattice at an extended time interval for both $J = 0.4$ and $0.6$.
The inset in (b) shows results for the $t$-$J$ and $t$-$J_z$ model  vs $tJ_\mathrm{eff}^{2/3}$ where $J_\mathrm{eff}=J$ for the $t$-$J$ model and $J_\mathrm{eff}=J/2$ for the $t$-$J_z$ model.
Short horizontal lines indicate values of $E_\mathrm{kin}$ in the respective ground states.}
\label{Fig:formation}
\end{figure}

In Fig.~\ref{Fig:formation} we present results of the hopping amplitude quench using three different models: $t$-$J$ model, Eq.~(1),  $t$-$J_z$ model, where in  Eq.~(1) only the $z$-components of spin operators are taken  into account,  and finally, the $t$-$J_z$  model on the Bethe lattice, Eq.~(2).
In the first two cases the functional space of Eq.~(\ref{lfs}) was used and in the case of the $t$-$J$ model, the initial wavefunction is given by Eq.~(\ref{def_polform}).
In all cases we observe ultrafast relaxation of  $E_\mathrm{kin}(t)$ in regime $tJ^{2/3}\lesssim 2$.
Comparison of relaxation dynamics after the phase quench, presented in Figs.~\ref{fig1}(b) and~\ref{fig1}(d), and the hopping amplitude quench, Fig.~\ref{Fig:formation}, shows very similar ultrafast relaxation followed by oscillations slightly above  their respective ground state values of $E_\mathrm{kin}$,  indicated  by short horizontal lines.
Again, the exception are results obtained on the Bethe lattice where $E_\mathrm{kin}(t)$ at long times oscillates much above its  value in the ground state, see also the inset of Fig.~\ref{Fig:formation}(b).
We therefore conclude that the characteristic time scale of the primary relaxation, $\tau \sim (\hbar/t_0)(J/t_0)^{-2/3}$ represents a general result, irrespective of the initial state.

\section{Quantitative comparison: $t$-$J$ model versus $t$-$J_z$ model} \label{s2}

We further make quantitative comparison of relaxation times on a square lattice for the $t$-$J$ model and the $t$-$J_z$ model.
A closer comparison  of $E_\mathrm{kin}(tJ^{2/3})$ in the ultrafast relaxation regime, {\it i.e.}, for times $tJ^{2/3}\lesssim 1$, reveals a somewhat faster relaxation in the $t$-$J$ than in the $t$-$J_z$ model.
While it is tempting to explain this difference with the lack of magnon excitations in the latter model, Fig.~\ref{fig2} undisputedly shows that in this short time interval the carrier releases its excess kinetic to local antiferromagnetic excitations.
Magnons thus in the ultrafast relaxation regime do not play a significant role.
We can make even a step further in the understanding of the slight discrepancy in the ultrafast relaxation between different models.
In the inset of Fig.~\ref{Fig:formation}(b) we present results of $E_\mathrm{kin}(tJ_\mathrm{eff}^{2/3})$ where $J_\mathrm{eff}=J$ for the $t$-$J$ model and $J_\mathrm{eff}=J/2$ for the $t$-$J_z$ model on a square lattice.
Note that different values of $J$, leading to $J_\mathrm{eff}$, were used only to rescale the time-axis.
We obtain nearly perfect overlap for $tJ_\mathrm{eff}^{2/3}\lesssim 1$.
To explain the  agreement  between different models  we note that the maximal gain of a high-energy ferromagnetic-like bond in the $t$-$J$ model is  twice larger than in the $t$-$J_z$ model. Relaxation is thus faster in the $t$-$J$ model because antiferromagnetic excitations are more efficient in absorbing  energy that in turn renders them more  effective in  slowing down the excited hole. 

\begin{figure}[!b]
\includegraphics[width=0.49\textwidth]{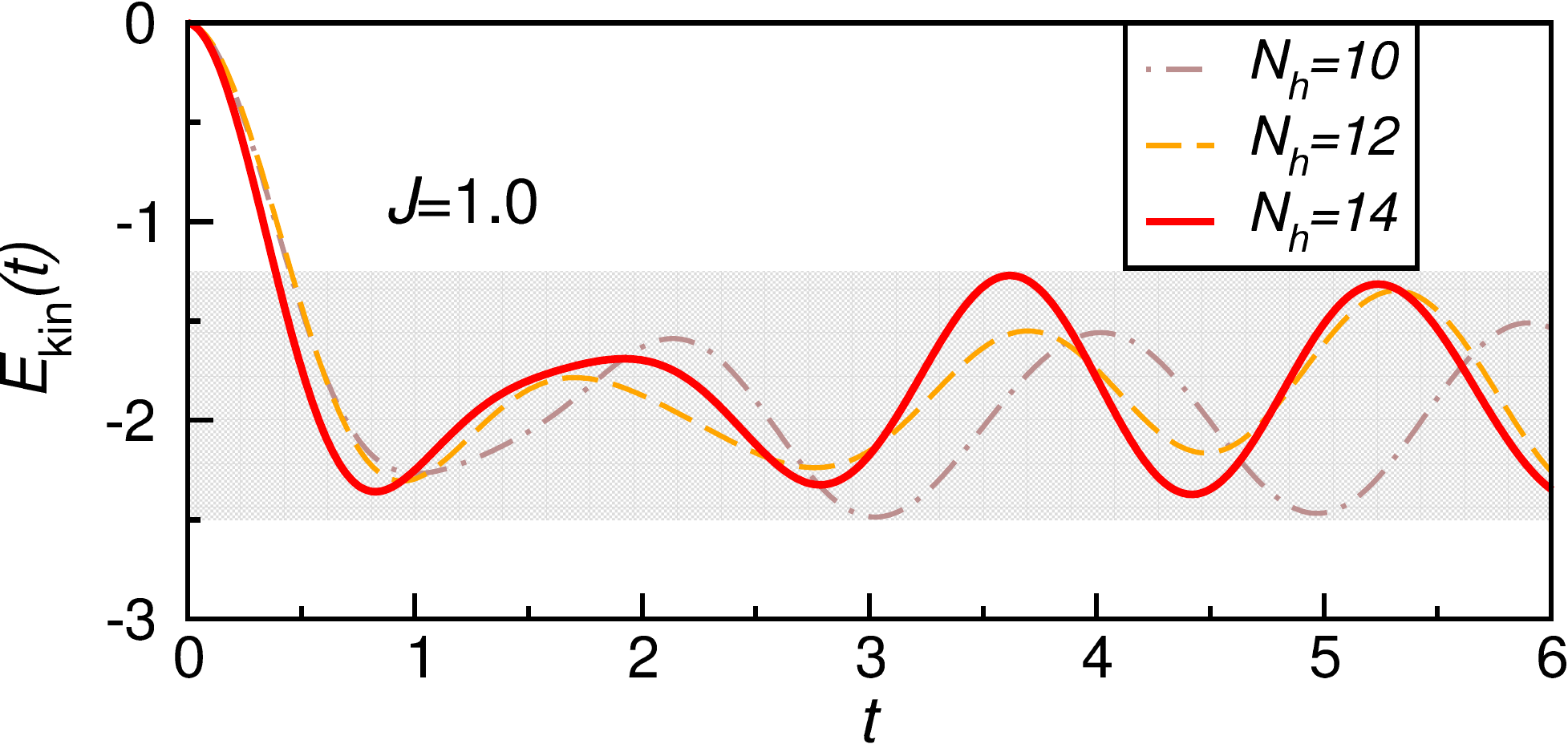}
\caption{
{\it Relaxation dynamics on a square lattice.}
Kinetic energy of the photo-carrier $E_\mathrm{kin}(t)$ for $J=1.0$. 
Curves represent results for different values of $N_h=10,12,14$ of the functional space generator defined in Eq.~(\ref{lfs}).
}
\label{Fig:sup2}
\end{figure}

\section{Oscillations after the primary relaxation} \label{s_oscillations}

After the initial fast relaxation of $E_\mathrm{kin}(t)$ presented in Fig.~\ref{fig1}, oscillations appear which are more pronounced at larger $J$.
While results for large $J$ should not be considered as experimentally relevant data, they are necessary for establishing the scaling of relaxation time with $J$.
In addition, the main focus of our study is devoted to the energy exchange between the photo-carrier and the antiferromagnetic background on the ultrafast time scale $\tau \sim (\hbar/t_0)(J/t_0)^{-2/3}$.
For the sake of convenience, we complete our analysis by studying finite-size effects of the oscillations of $E_\mathrm{kin}(t)$.

Figure~\ref{Fig:sup2} shows results for the kinetic energy $E_\mathrm{kin}(t)$ in the $t$-$J$ model on a square lattice at $J=1.0$, for different sizes of the functional space.
Shortly after the primary relaxation, oscillations emerge with amplitudes that do not decrease when $N_h$ increases.
It therefore excludes the possibility that the oscillations on the short timescales are numerical artefacts originating from the truncation of the Hilbert space.
We expect that in the secondary stage of the relaxation, these oscillations decay due to propagation of antiferromagnetic excitations.
Beside the simple analytical scenario addressed towards the end of the paper, we refrain from making explicit claims concerning quantitative values of this secondary relaxation time since our numerical method does not allow precise enough time evolution in this long-time regime.

Our results suggest the oscillations represent an inherent property of the $t$-$J$ model which, however, occurs only for unrealistically large $J$ and on the short timescales.
In the regime of large $J\sim t_0$, the photo-carrier induces only {\it short} string states.
The resulting oscillations hence emerge due to the transitions between a few lowest-energy string states involved in the relaxation, and scale as well with $J^{2/3}$.

\bibliography{references.bib}
\end{document}